\documentclass[prl,twocolumn,showpacs,preprintnumbers,aps,nofootinbib]{revtex4-1}

\usepackage{graphicx}
\usepackage{dcolumn}
\usepackage{bm}
\usepackage{amsmath,amssymb,amsfonts}
\usepackage{latexsym}

\usepackage{color}

\def\nn    {\nonumber}

\def\bsg{B\to X_s \gamma}

\def\rtt{\rho_{tt}}

\def\fbi{fb$^{-1}$\;} 
\def\bhpm{$cg \to bH^+$}
\def\cgbhpm{$cg \to b H^+ \to b t\bar b$\;}
\begin{document}


\title{\boldmath
Sub-TeV $H^+$ Boson Production as Probe of  Extra Top Yukawa Couplings}

\author{Dilip Kumar Ghosh,$^1$ Wei-Shu Hou$^2$ and Tanmoy Modak$^2$}
\affiliation{$^1$School of Physical Sciences, Indian Association for the Cultivation of Science, Kolkata 700032, India}
\author{}
\affiliation{$^2$Department of Physics, National Taiwan University, Taipei 10617, Taiwan}


\begin{abstract} 
We suggest searching for the charged Higgs boson 
at the Large Hadron Collider (LHC) via $cg \to b H^+ \to b t \bar b$. 
In the general two Higgs Doublet Model, 
extra top Yukawa couplings $\rho_{tc}$ and $\rho_{tt}$ 
can drive the disappearance of antimatter from the Universe,
while $\bar cbH^+$ and $\bar tbH^+$ couple with strength 
$\rho_{tc}V_{tb}$ and $\rho_{tt}V_{tb}$, respectively. 
For $\rho_{tc},\, \rho_{tt} \sim 0.5$ and $m_{H^+} \sim 300$--500 GeV, 
evidence could emerge from LHC Run 2 data at hand, 
and discovery by adding Run 3 data in the near future.
\end{abstract}

\maketitle

\begin{figure}[b]
\center
\includegraphics[width=0.31\textwidth]{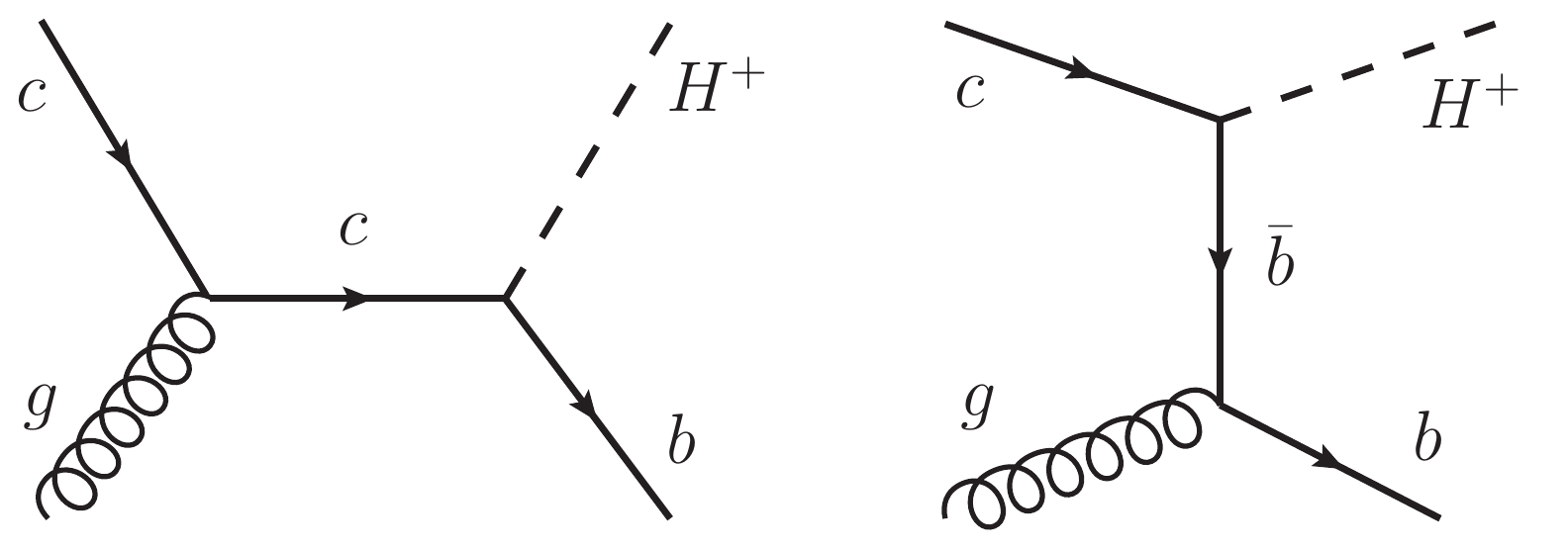}
\caption{Feynman diagrams for $cg\to b H^+$.}
\label{feyndiag}
\end{figure}

\noindent{\it Introduction.---}
The discovery of
the Higgs boson $h(125)$ 
at the LHC~\cite{h125_discovery} 
 {suggests} a weak scalar doublet, but there is 
no principle that precludes the existence of a second doublet. 
%
%
Having two Higgs doublets (2HDM){, one has a} 
{charged $H^+$ boson
 {plus} the $CP$-even/odd scalar bosons $H$, $A$~\cite{Branco:2011iw}.
}
We propose a novel process, $cg \to bH^+$ (see Fig.~\ref{feyndiag})
followed by $H^+ \to t\bar b$, that may lead to
the discovery of {the exotic} $H^+$ boson in the near future. 


In the popular 2HDM type~II (2HDM-II), 
up- and down-type quark masses arise from separate doublets~\cite{Branco:2011iw},
hence mass and Yukawa matrices are 
simultaneously diagonalized, just like in {the Standard Model (SM)}.
The model motivates {an} $H^+$ search at the LHC 
via the process $\bar bg \to \bar tH^+$~\cite{Aaboud:2018cwk,Sirunyan:2019arl} 
which goes through the $\bar t bH^+$ coupling, while
%
the $cg \to bH^+$ process is suppressed 
by the Cabibbo-Kobayashi-Maskawa (CKM) matrix element ratio 
$|V_{cb}/V_{tb}|^2 \sim 1.6 \times 10^{-3}$.
But in the general 2HDM (g2HDM) with {\it extra} 
Yukawa couplings~\cite{Chen:2013qta},
$\bar cbH^+$ and $\bar tbH^+$ couple with strength
$\rho_{tc}V_{tb}$ and $\rho_{tt}V_{tb}$, respectively,
and $cg \to bH^+$ is not CKM-suppressed.

%
{The extra top Yukawa couplings}~\cite{Chen:2013qta} $\rho_{tc}$ and $\rho_{tt}$
{are not well constrained.}
{If both are ${\cal O}(1)$, i.e.} 
the top Yukawa coupling strength {$\lambda_t$} in SM, 
they facilitate the production and decay in 
$cg \to bH^+ \to bt\bar b$~{\cite{Iguro:2017ysu, Gori:2017tvg}}, 
with the signature of lepton plus missing energy and three $b$-jets.
It is known~\cite{Fuyuto:2017ewj} that $\rho_{tc}$ and $\rho_{tt}$ 
at ${\cal O}(1)$ can {\it each} drive electroweak baryogenesis (EWBG), 
hence account for 
the disappearance of antimatter shortly after the Big Bang, 
one of the biggest mysteries.
Perhaps equally interesting, 
when the ACME 2018 bound~\cite{Andreev:2018ayy} 
on electron electric dipole moment (eEDM) seemed to rule out
the $\rho_{tt}$ parameter space of Ref.~\cite{Fuyuto:2017ewj},
a second paper~\cite{Fuyuto:2019svr} brought in 
the extra electron Yukawa coupling, $\rho_{ee}$,
and showed that a natural cancellation mechanism 
can survive the ACME18 bound, and with 
expanded parameter space for EWBG.
This gives strong motivation for the $cg \to bH^+ \to bt\bar b$ search.
{
The recent CMS hint of an ``excess''~\cite{Sirunyan:2019wph} 
in $gg \to A \to t\bar t$  at $m_A \sim 400$~GeV 
 could also arise from $\rho_{tt} \sim {\cal O}(1)$~\cite{Hou:2019gpn}.}

In this Letter, we first show that {the $H$, $A$ and $H^+$} bosons in g2HDM 
can be sub-TeV in mass while satisfying all known constraints.
This is in contrast with the absence of beyond SM (BSM) signatures 
so far at the LHC, with bounds often reaching multi-TeV in scale.
We then show that $\rho_{tc}$, $\rho_{tt}$ at ${\cal O}(1)$ 
is allowed by current $\bar bg \to \bar tH^+$ and other search bounds. 
Full Run~2 data 
could already give evidence for $cg \to bH^+ \to bt\bar b$, 
and discovery is possible by adding Run~3 data.

\vskip0.08cm
\noindent{\it Dimension-4 Higgs Couplings.---}
Besides gauge couplings, Higgs bosons uniquely 
possess two additional sets of dimension-4 couplings:
Higgs quartic and Yukawa interactions.
In the Higgs basis, one can write the most general $CP$-conserving 
potential~\cite{Davidson:2005cw,Hou:2017hiw} in g2HDM as
\begin{align}
 & V(\Phi,\Phi') = \mu_{11}^2|\Phi|^2 + \mu_{22}^2|\Phi'|^2
                              - (\mu_{12}^2\Phi^\dagger\Phi' + h.c.)  \nn\\
 & \quad + {1\over 2}\eta_1|\Phi|^4 + \frac{1}{2}\eta_2|\Phi'|^4 + \eta_3|\Phi|^2|\Phi'|^2
              + \eta_4 |\Phi^\dagger\Phi'|^2\nn\\
 & \;\; + \bigg[\frac{1}{2}\eta_5(\Phi^\dagger\Phi')^2
     + \left(\eta_6 |\Phi|^2 + \eta_7|\Phi'|^2\right) \Phi^\dagger\Phi' + h.c.\bigg],
\label{pot}
\end{align}
where all quartic couplings $\eta_i$ are real, 
$\Phi$ induces spontaneous symmetry breaking by 
the vacuum expectation value $v$, i.e. $\mu_{11}^2 = - \frac{1}{2}\eta_1 v^2 < 0$, 
while $\left\langle \Phi'\right\rangle =0$ hence $\mu_{22}^2 > 0$. 
The minimization condition 
$\mu_{12}^2 = \frac{1}{2}\eta_6 v^2$ 
reduces the parameter count to nine. 
%
From Eq.~(\ref{pot}) one finds $m_{h^{(0)}}^2 = \eta_1v^2$,
%
 $m_{H^\pm}^2 
   = \mu_{22}^2 + \frac{1}{2}\eta_3 v^2$
and
 $m_{H^{(0)}, A}^2 
   =m_{H^+}^2 + \frac{1}{2}(\eta_4 \pm \eta_5) v^2$. 
%
Finally, $\eta_6$ mixes $h^{(0)}$ and $H^{(0)}$ 
into $h$ and $H$.
%
{T}he emergent alignment phenomenon, that
 $h$ resembles {the Higgs boson of the SM} 
so well~{\cite{Khachatryan:2016vau,Sirunyan:2018koj,Aad:2019mbh},
implies that}
the $h$--$H$ mixing angle $c_\gamma \equiv \cos\gamma$
{({denoted usually as} $-\cos(\beta - \alpha)$) 
} 
is rather small.

The 
Yukawa couplings to quarks are~\cite{Davidson:2005cw, Hou:2017hiw}
\begin{align}
\mathcal{L} = 
 - & \frac{1}{\sqrt{2}} \sum_{f = u, d} 
 \bar f_{i} \Big[\big(-\lambda^f_i \delta_{ij} s_\gamma + \rho^f_{ij} c_\gamma\big) h \nn\\
 & + \big(\lambda^f_i \delta_{ij} c_\gamma + \rho^f_{ij} s_\gamma\big)H
    - i\,{\rm sgn}(Q_f) \rho^f_{ij} A\Big]  R\, f_{j} \nn\\
 - & \bar{u}_i\left[(V\rho^d)_{ij} R-(\rho^{u\dagger}V)_{ij} L\right]d_j H^+ 
 +{h.c.},
\label{eff}
\end{align}
where $\lambda_{i}^f  = {\sqrt{2}\,m_i^f}/{v}$, 
$L, R = (1\mp \gamma_5)/2$ and $s_\gamma \equiv \sin\gamma$.
Note that the $A$, $H^+$ couplings are independent of $c_\gamma$, 
{while} in the alignment limit of $c_\gamma \to 0$, $h$ couples diagonally 
{and} $H$ carries the extra Yukawa couplings $\rho_{ij}^f$.
Thus, besides mass-mixing hierarchy protection~{\cite{Hou:1991un,Chang:1993kw,Atwood:1996vj}} 
of flavor changing neutral Higgs (FCNH) couplings,
alignment provides~\cite{Hou:2017hiw} further safeguard,
without the need of Natural Flavor Conservation~\cite{Glashow:1976nt}.
The importance of $\rho_{tt}$ and $\rho_{tc}$ was emphasized~\cite{Chen:2013qta}
already at the $h(125)$ discovery, and was subsequently 
shown~\cite{Fuyuto:2017ewj} to possibly drive EWBG.

\begin{figure}[t]
\center
\includegraphics[width=0.389 \textwidth]{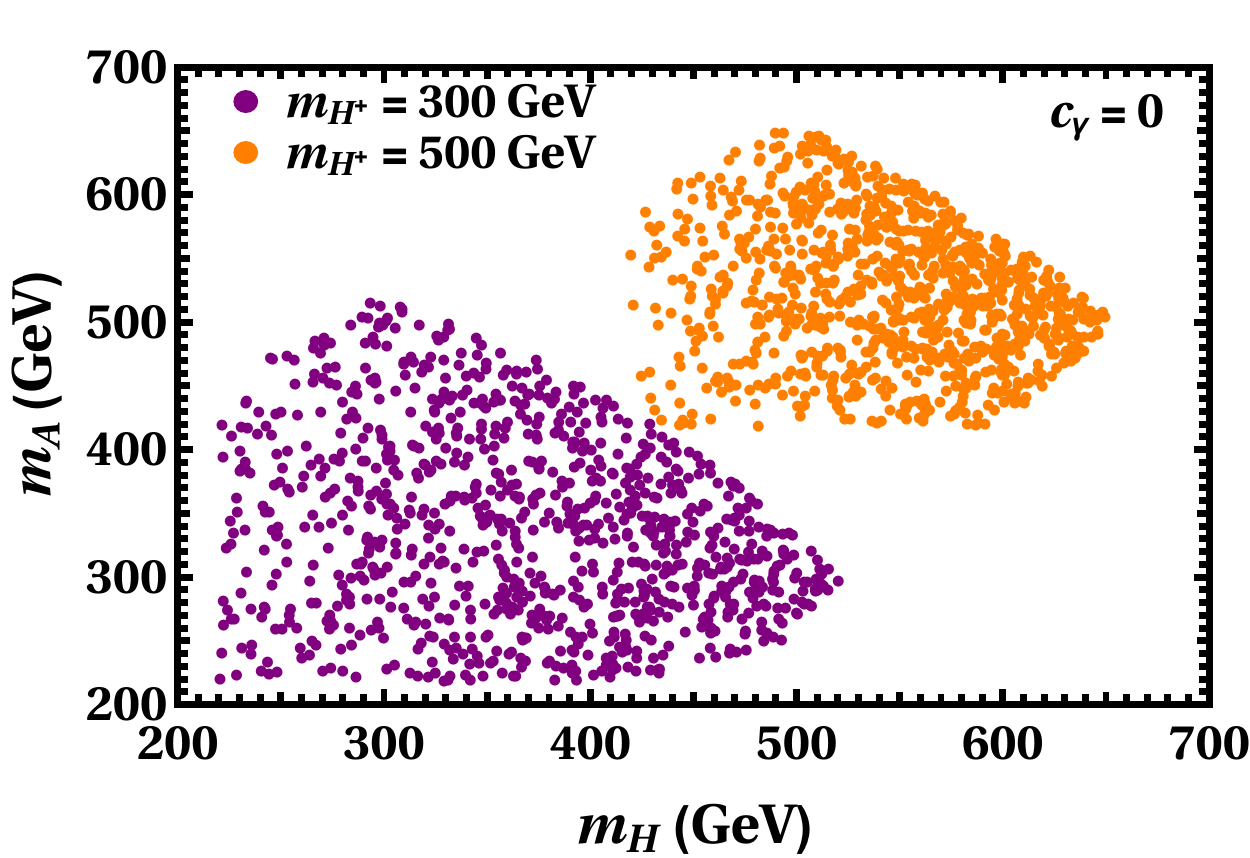}
\caption{
Scan points in $m_H$--$m_A$ plane that pass 
positivity, perturbativity, unitarity and 
oblique parameter constraints. 
}
\label{massscan}
\end{figure}

From Eq.~(2) one finds that the leading $\bar cbH^+$ and $\bar tbH^+$
couplings are $\rho_{tc} V_{tb}$ and $\rho_{tt} V_{tb}$, 
respectively~{\cite{rho-d}},
where there is no CKM-suppression of the former~{\cite{cbH+}}
as in 2HDM-II.
In this Letter, we take $m_{H^+} > m_t$~{\cite{tcbH+}}
and focus on the \cgbhpm process at the LHC. 
%
{
Note that the $gg \to \bar cbH^+$ process discussed in Ref.~\cite{Gori:2017tvg} 
bears some similarity, 
but Fig.~1(left) was not mentioned explicitly, 
and a detailed collider study was not performed, 
hence the promise was not sufficiently demonstrated.
}


\vskip0.08cm
%
\noindent{\it Constraints on Higgs Parameters.---}
Higgs quartics 
need to satisfy positivity, perturbativity and tree-level unitarity, 
which we implement via 2HDMC~\cite{Eriksson:2009ws}. 
We express~\cite{Davidson:2005cw,Hou:2017hiw} 
$\eta_1$, $\eta_{3{\rm -}6}$ in terms of 
$\mu_{22}$, $m_{h,\, H,\, A,\, H^+}$
 (all normalized to $v$) and $\cos\gamma$, 
plus $\eta_2$, $\eta_7$ that do not enter Higgs masses.
%
%
{Since $H^+$ Yukawa couplings do not depend on $c_\gamma$,
which is known to be small,}
we set 
$c_\gamma = 0$ {for simplicity} while fixing $m_h \cong $ 125 GeV,
hence~\cite{Hou:2017hiw} $\eta_6 = 0$ and $\eta_1 = m_h^2/v^2$.
{Thus,} e.g. $t \to ch$ does not constrain $\rho_{tc}$.
%
%
{In the common Higgs basis, 
we identify  $\eta_{1-7}$ with the input parameters $\Lambda_{1-7}$ to 2HDMC.
 }

For fixed $m_{H^+}$, 
we randomly generate the parameters in the ranges 
{$|\eta_{2-5,\,7}| \leq 3$ (positivity requires $\eta_2 > 0$),} 
$\mu_{22} \in [0, 1]$  TeV,
and $m_{A,\, H} \in [m_{H^+}-m_W, 650$~GeV]
to forbid $H^+ \to AW^+, HW^+$. 
%
%
{We then use 2HDMC for scanning, 
 where the electroweak oblique parameter constraints
{(including correlations)} are imposed, 
e.g. the $2\sigma$ range of $-0.17 < T < 0.35$}~\cite{T-param},
%
which restricts~\cite{Froggatt:1991qw,Haber:2015pua} 
the scalar masses 
hence the $\eta_i$s. 
%
{Scan points satisfying these constraints}
are plotted in Fig.~\ref{massscan} in the $m_H$--$m_A$ plane
 for $m_{H^+} = 300$, 500 GeV,
illustrating that finite parameter space exist.
We choose a benchmark for each $m_{H^+}$ value  
and list the parameters in Table~\ref{bench}.
More details of {our} scanning procedure {is given} in Ref.~\cite{Hou:2019qqi}.

\begin{figure*}[t!]
\center
\includegraphics[width=0.33 \textwidth]{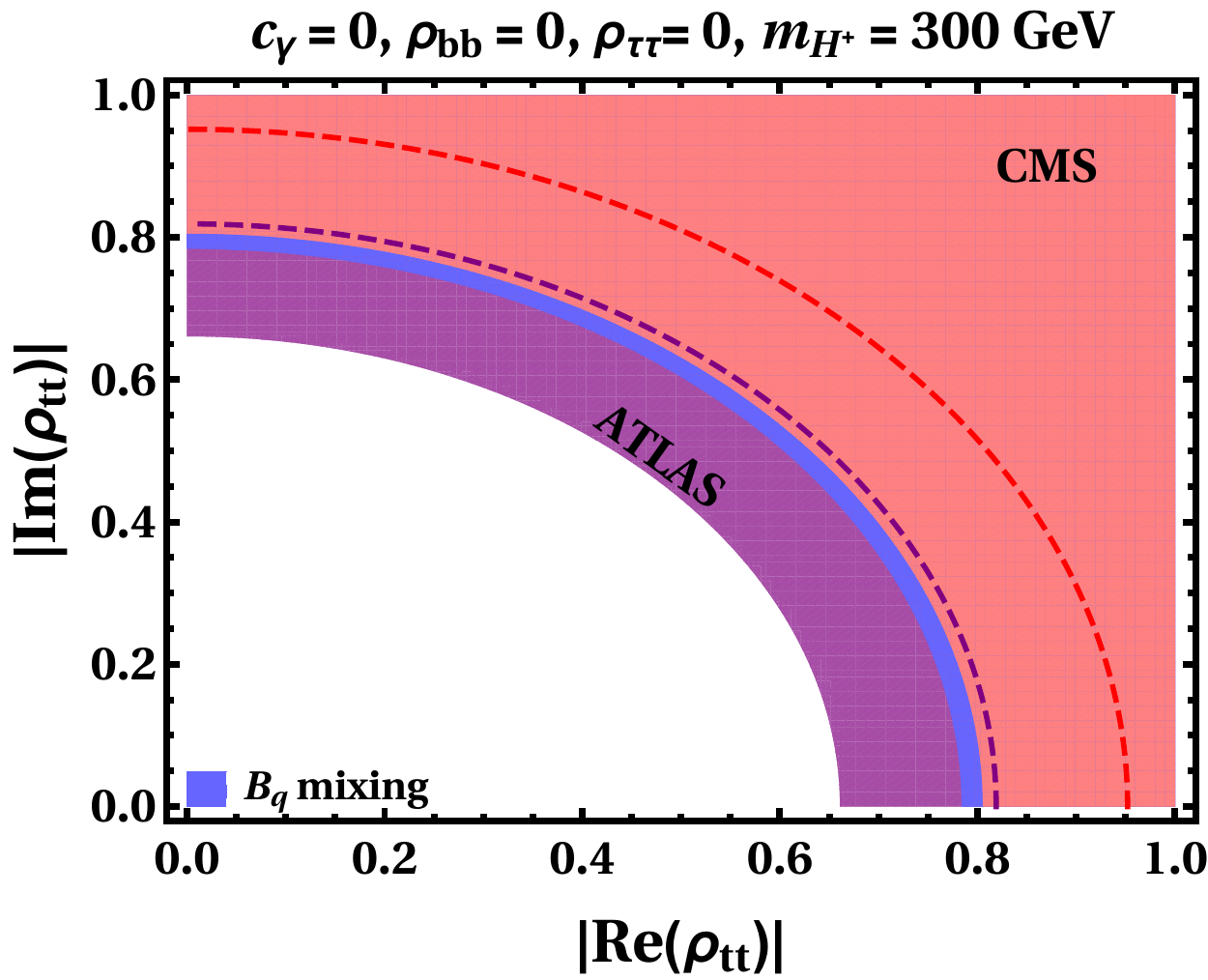}
\includegraphics[width=0.33 \textwidth]{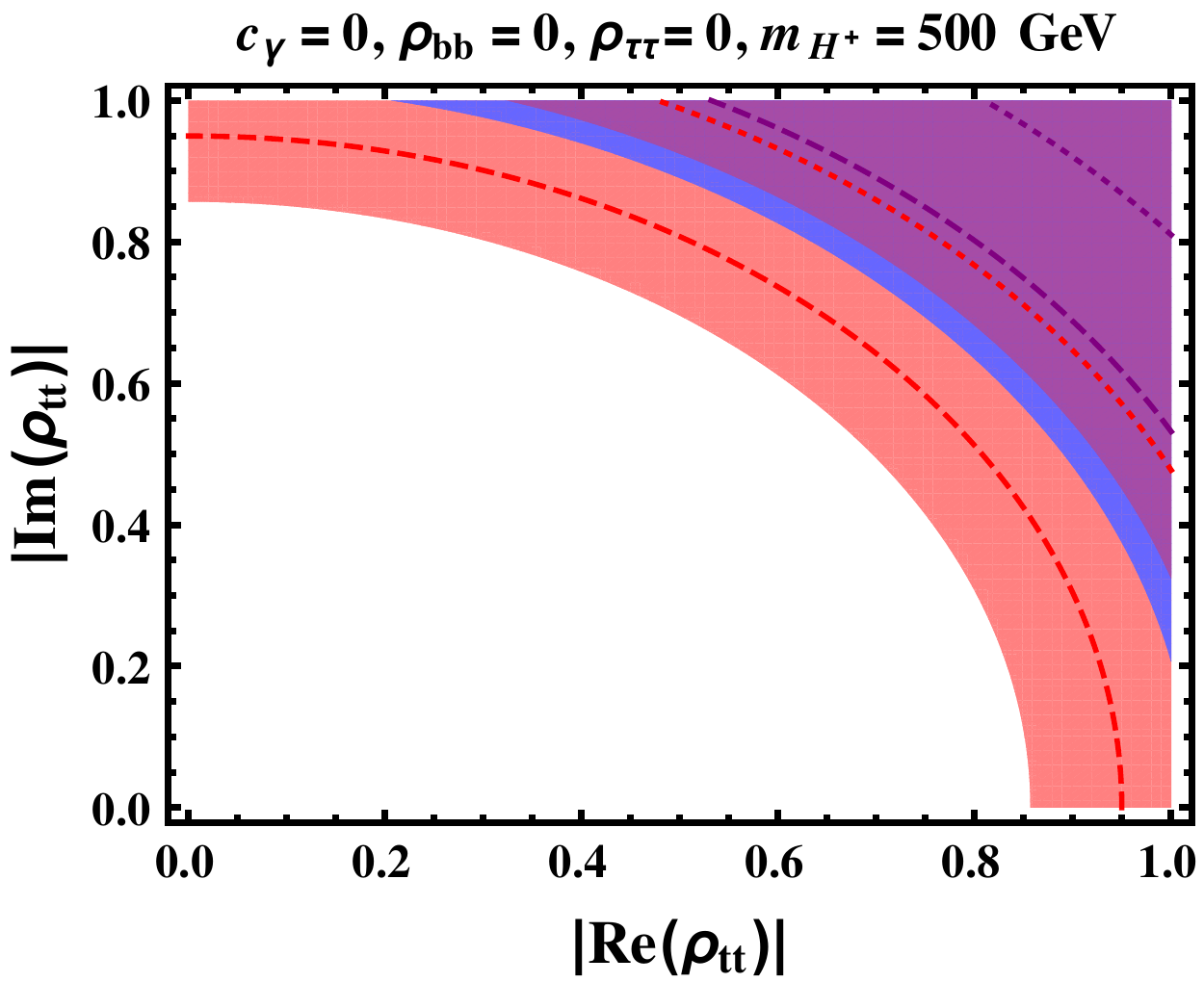}
\caption{
Constraint from $B_q$ mixings on $\rtt$
 (blue shaded region with area to upper right excluded)
 assuming all other $\rho_{ij}=0$.  
The excluded regions from $bg \to \bar t (b) H^+ \to \bar t (b) t \bar b$ searches 
by ATLAS~\cite{Aaboud:2018cwk} and CMS~\cite{Sirunyan:2019arl} 
for $\rho_{tc} = 0$ are overlaid (purple and red shaded),
which is weakened for $\rho_{tc}=0.4$ (dash) 
{and 0.8 (dots). See text for details}.
}
\label{rhottconst}
\end{figure*}

\vskip0.08cm
%
\noindent{\it Flavor Constraints.---} 
%
Flavor constraints on $\rho_{tt}$ and $\rho_{tc}$ 
are not particularly strong~\cite{Chen:2013qta,Altunkaynak:2015twa}. 
For $m_{H^+} \lesssim 500$ GeV, $B_q$ 
mixings ($q$ = $d$, $s$) provide 
the most stringent constraint. 
{An $H^+$ effect from $\rho_{ct}$ to the $M^q_{12}$ amplitude
 is enhanced by} $|V_{cq}/ V_{tq}|\sim 25$, 
hence $\rho_{ct}$ must be turned off~\cite{Altunkaynak:2015twa}.
Assuming all $\rho_{ij}$ vanish except $\rho_{tt}$,
we have $M^q_{12}/M^{q}_{12}|^{\rm{SM}}= C_{B_q}$, with negligible phase.
Allowing $2\sigma$ error on 
$C_{B_d} = 1.05\pm 0.11$ and $C_{B_s} = 1.11\pm0.09$~\cite{utfitrse}, 
we find the blue shaded exclusion region 
(extending to upper-right) in Fig.~\ref{rhottconst},  
where the left (right) panel is for BP1 (BP2).
 {The constraint from $H^+$ effects via charm loops~\cite{Crivellin:2013wna} 
 gives $\rho_{tc} \lesssim 1\, (1.7)$ for BP1 (BP2).
 }

\begin{table}[t]
\centering
{
\begin{tabular}{c |c| c| c| c | c | c| c | c |c} 
\hline
 & 
 $\eta_2$ & $\eta_3$ & $\eta_4$ & $\eta_5$ 
 & $\eta_7$
 & \,${\mu_{22}^2\over v^2}$\, &  $m_{H^+}$  &  $m_A$  &  $m_H$  \\
\hline\hline
\,BP1\, & 
  1.40 & 0.62 & 0.53 & \,\;\ 1.06 & 
 {$-0.79$} & 1.18 & 300 & 272 & 372 \\
\,BP2\, & 
 0.71 & 0.69 & 1.52 & {$-0.93$} & 
 {\,\;\ 0.24} & 3.78 & 500  & 569   & 517 \\
\hline
\hline
\end{tabular}
}
\caption{
Benchmark points BP1 and BP2, 
withg $\eta_6 = 0$ hence $\eta_1 \cong 0.258$. 
Higgs masses are in GeV.}
\label{bench}
\end{table}

$\bsg$ {puts} a strong constraint 
on $m_{H^+}$ in 2HDM-II, but weakens for g2HDM
{due to extra Yukawa couplings}.
In fact, an $m_t/m_b$ enhancement factor 
constrains $\rho_{bb}$ more strongly~\cite{Altunkaynak:2015twa} 
than $\rho_{tt}$. 
Taking $\rho_{bb}$ as small, 
the constraint on $\rtt$ 
falls outside the range of Fig.~\ref{rhottconst}.
The $\bsg$ 
constraint on $\rho_{tc}$ via charm loop
is weaker than $B_q$ mixing~\cite{Altunkaynak:2015twa}.
{Note that flavor constraints would grow weaker
 for $m_{H^+}$ heavier than our benchmarks.
 }
%

\vskip0.08cm
%
\noindent{\it Collider Constraints.---}
%
{To focus on our signal process}, we set all $\rho_{ij} = 0$ 
 except $\rho_{tt}$ and $\rho_{tc}$ for simplicity. 

For finite $\rho_{tt}$, one can have $\bar bg \to \bar t (b) H^+$~\cite{hpmtb} 
followed by $H^+\to t \bar b$ (charge conjugate process implied). 
Searches 
at $13$ TeV provide model independent bounds on
$\sigma(pp\to \bar t {(b)} H^+) \,  \mathcal{B}(H^+  \to t \bar b)$, 
for $m_{H^+}=200$ GeV to  2 (3)~TeV
 for ATLAS~\cite{Aaboud:2018cwk} (CMS~\cite{Sirunyan:2019arl}). 
%
%
Using the Monte Carlo (MC) event generator
MadGraph5\_aMC@NLO~\cite{Alwall:2014hca} with 
default NN23LO1 parton distribution function (PDF)~\cite{Ball:2013hta} 
and effective model implemented in FeynRules~2.0~\cite{Alloul:2013bka}, 
we calculate $\sigma(pp\to \bar t {(b)} H^+)\, (H^+  \to t \bar b)$ 
at leading order (LO) for a reference $|\rho_{tt}|$,
then rescale by $|\rtt|^2\,\mathcal{B}(H^+ \to t \bar b)$ to get the upper limits. 
For $m_{H^+} = 300$, 500 GeV and 
with $\rho_{tc}=0$ (hence $\mathcal{B}(H^+ \to t \bar b) \sim 100\%$),
we plot the extracted 
ATLAS (CMS) 95\% C.L. bounds on $\rtt$ as
the red (purple) shaded regions in Fig.~\ref{rhottconst}.
%
%
The ATLAS/CMS limit is more/less stringent than 
$B_q$ mixing for BP1 ($m_{H^+} = 300$ GeV),
while opposite for BP2 ($m_{H^+} = 500$ GeV).
The exclusion bands are overlaid to illustrate this.

%
Heavy Higgs searches via $gg\to H/A \to t \bar t$ 
can constrain $\rho_{tt}$. 
ATLAS~\cite{Aaboud:2017hnm} searched 
at 8 TeV for $m_{A/H} > 500$~GeV;  
with $36$ fb$^{-1}$~at $13$ TeV, CMS constrains 
the ``coupling modifier''~\cite{Sirunyan:2019wph}
for $m_{A/H} = $ 400--750 GeV for various  
$\Gamma_{A/H}/m_{A/H}$ 
 values. 
Both ranges are above BP1, 
while for BP2 the bounds are weaker than
results shown in Fig.~\ref{rhottconst}(right).
{The} CMS ``excess'' at $m_A \sim 400$ GeV~\cite{Sirunyan:2019wph}
{is discussed} later.

Based on 137~\fbi at 13 TeV, the CMS $4t$ search~\cite{Sirunyan:2019wxt} 
constrains $\rho_{tc}$ and $\rho_{tt}$. 
{We first note that the direct limits from} 
$\sigma(pp \to t \bar t A/t \bar t H)\,\mathcal{B}(A/H \to t \bar t)$ 
for $m_{A/H}\in [350,\, 650]$ GeV 
%
%
are again weaker {than results} shown Fig.~\ref{rhottconst}.
With both $\rho_{tc}$ and $\rho_{tt}$ finite, 
the $cg\to t H/tA \to t t \bar t$ process~\cite{Kohda:2017fkn} 
{can feed the Signal Region,} SR12, of the CMS $4t$ search
if all three top quarks decay semileptonically.
{As $cg \to tH/tA \to tt\bar t$ barely occurs for BP1 
because of low $m_{A,\, H}$ values, 
this applies only to BP2.}
SR12 requires~\cite{Sirunyan:2019wxt} at least three leptons, 
four jets with at least three $b$-tagged, 
plus missing $p_T$. 
%
%
%
{Following Ref.~\cite{Hou:2019gpn}, 
we generate events 
and interface with PYTHIA~6.4~\cite{Sjostrand:2006za} 
for showering and hadronization, 
adopt MLM merging~\cite{Mangano:2006rw} 
 of matrix element and parton shower, 
then feed into Delphes~3.4.2~\cite{deFavereau:2013fsa} 
for CMS-based fast detector simulation, 
including $b$-tagging and $c$- and light-jet rejection.
We find $\rho_{tt} \gtrsim 1$ is excluded if $\rho_{tc} \sim 0.8$
for BP2.}
{Noting that $H^+ \to c\bar b$ decay from finite $\rho_{tc}$
would dilute ${\cal B}(H^+ \to t\bar b)$ and 
soften the $bg \to \bar t (b) H^+$ constraint, 
we illustrate this effect by the dash (dot) curves
in Fig.~\ref{rhottconst}(right) for $\rho_{tc} = 0.4\, (0.8)$.}
%
%

{The $cg \to tH/tA \to tt\bar c$ process~\cite{Kohda:2017fkn} 
can feed the Control Region for $t\bar tW$ (CRW) background}
of CMS 4$t$ study when both tops decay semileptonically. 
With CRW defined by same-sign dileptons ($e$ or $\mu$), 
$p_T^{\rm miss}$, and up to five jets with at least two $b$-tagged,
%
%
%
we follow Refs.~\cite{Hou:2018zmg,Hou:2019gpn} 
%
%
%
{and find $\rho_{tc}\gtrsim 0.4$ is excluded for BP1, 
which is stronger than the $B_q$ mixing bound, and
with little dependence on $\rho_{tt}$.
For BP2, we find that CRW gives comparable limit as SR12.
Thus, we illustrate in Fig.~\ref{rhottconst}(left) 
the softened $bg \to \bar t (b) H^+$ constraint 
only for $\rho_{tc} = 0.4$.
}

We remark in passing that the ATLAS search for 
same-sign di-leptons and $b$-jets~\cite{Aaboud:2018xpj}, 
%
or {search for supersymmetry} in similar event topologies~{\cite{Aad:2019ftg}},
impose stronger cuts and in general do not give relevant constraints.

\vskip0.08cm
%
\noindent{\it Collider Signature for $cg \to bH^+ \to bt\bar b$.---}
%
We now show that the $cg \to bH^+ \to bt\bar b$ process,
or $pp \to b H^+ + X \to b t \bar b +X$, is quite promising.

For illustration, we {\it conservatively} take 
$|\rho_{tc}|=0.4$, $|\rho_{tt}|= 0.6$ for both BPs. 
%
{Receiving no CKM suppression, the approximate} 
$H^+ \to c\bar b,\; t\bar b$ branching ratios 
are 50\%, 50\% for BP1, and 36\%, 64\% for BP2.
%
%
Assuming $t\to b \ell \nu_\ell$ ($\ell =e,\,\mu$),
the signature is one charged lepton, $p^{\rm miss}_T$, and three $b$-jets. 
Subdominant contributions such as PDF-suppressed   
$bg\to \bar c H^+, \bar t H^+ \to \bar c t \bar b, \bar t c \bar b$
with $c$-jet mistagged as $b$-jet, 
and $\rho_{tt}$-induced $bg\to \bar t H^+ \to \bar t t \bar b$ 
with one top decaying hadronically, are included as signal.
%
There is also $c\bar b \to H^+ \to c \bar b, t \bar b$~{\cite{H+incl}}, 
but these suffer from QCD and top backgrounds. 
The dominant backgrounds for \bhpm  arise from 
$t\bar t+$jets, $t$- and $s$-channel single-top ($tj$), $Wt+$jets, 
with subdominant backgrounds from $t\bar t h$ and $t\bar t Z$. 
Minor contributions from Drell-Yan and $W+$jets, $4t$, $t\bar t W$, $tWh$ 
are {combined} under ``other''.

Signal and background samples are generated at LO 
for $14$ TeV as before by MadGraph, interfaced with PYTHIA 
%
and fed into Delphes for fast detector simulation adopting 
default ATLAS-based detector card. 
%
The LO $t\bar t +$jets background is normalized to
NNLO by a factor $1.84$~\cite{twiki}, and 
factors of 1.2 and 1.47~\cite{twikisingtop} for
$t$- and $s$-channel single-top. 
The LO $Wt+$jets background is normalized to NLO 
by a factor 1.35~\cite{Kidonakis:2010ux}, 
whereas the subdominant $t\bar t h$, $t\bar t Z$ receive 
factors of 1.27~\cite{twikittbarh}, 1.56~\cite{Campbell:2013yla}.
The DY+jets background is normalized to NNLO 
by a factor 1.27~\cite{Li:2012wna}. 
Finally, the $4t$ and $t\bar t W^-$ ($t\bar t W^+$) cross sections at LO 
are adjusted to NLO by factors of 2.04~\cite{Alwall:2014hca} 
and 1.35 (1.27)~\cite{Campbell:2012dh}.
The $tWh$ and $W+$jets backgrounds are kept at LO.
Correction factors for other charge conjugate processes are assumed to be the same, 
and the signal cross sections are kept at LO.

Events are selected with one lepton, 
at least three jets with three $b$-tagged, and $E^{\rm miss}_T  >35$ GeV. 
Jets are reconstructed by anti-$k_t$ algorithm using radius parameter $R=0.6$. 
The lepton $p_T$ should be $> 30$ GeV, 
with all three $b$-jet $p_T >20$ GeV, 
and pseudo-rapidity ($|\eta|$) of lepton and $b$-jets $<2.5$. 
The $\Delta R$ separation between a $b$-jet and the lepton, 
or any $b$-jet pair, should be $> 0.4$. 
The sum of the lepton and three leading $b$-jet transverse momenta $H_T$ 
should be $>350$ (400) GeV for BP1 (BP2). 
%
%
We have not optimized the selection cuts for 
$H_T$, $p_T$, $E^{\rm miss}_T$, etc. 
The total background cross section~B$_{\rm tot}$ after selection cuts 
and its various components, together with the signal cross section Sig, 
are given in Table~\ref{bkgcomptbhpm}.

\begin{table}[t]
\centering
\begin{tabular}{c |c| c|c| c | c| c| c |c}
\hline
  & $t\bar tj$s & $tj$ & $Wtj$s & $t\bar t h$ & $t\bar t Z$ & other
  & B$_{\rm tot}$ & Sig  \\
\hline
\hline
 \,BP1\, & \,1546\, & \,42\, & 27 & \,4.2\, & 1.5 & 3.1 & \,1627\, & \,11.4\,  \\
 \,BP2\, & \, 1000 \, & 27 & 16 & 2.9 & 1.2 & 1.9 &\  1049\, & \,9.3\,   \\
\hline
\hline
\end{tabular}
\caption{
Background and signal (Sig, for $\rho_{tc} = 0.4$, $\rho_{tt} = 0.6$) 
cross sections (in fb) at $14$ TeV 
after selection cuts.
}
\label{bkgcomptbhpm}
\end{table}

We estimate the statistical significance from Table~\ref{bkgcomptbhpm} 
with $\mathcal{Z} = \sqrt{2[ (S+B)\ln( 1+S/B )-S ]}$~\cite{Cowan:2010js}.
For 137, 300 and 600 fb$^{-1}$, the significance for \bhpm\ is at
$\sim 3.3\sigma$, $4.9\sigma$, $6.9\sigma$ 
($\sim 3.4\sigma$, $5.0\sigma$, $7.1\sigma$) for BP1 (BP2).
{{Reanalyzing} the signal and backgrounds 
for 13 TeV 
at $137$ fb$^{-1}$, we find similar significance.}
Thus, full Run~2 data could already show evidence, 
{and combining ATLAS and CMS data is encouraged.}
Discovery is possible with combined Run~2 and 3 data. 


\vskip0.08cm
\noindent{\it Discussion and Summary.---}
A $3.5\sigma$ local ($1.9\sigma$ global) excess at $m_A\approx 400$ GeV 
was reported by CMS~\cite{Sirunyan:2019wph} in $gg\to A \to t \bar t$ search. 
%
%
{The excess can be explained~\cite{Hou:2019gpn} with sizable
$\rho_{tt}\sim 1.1$,  $\rho_{tc}\sim 0.9$ 
for $m_{H} \gtrsim\,500$ GeV, $m_{H^+} \gtrsim 530$ GeV. 
The bound on $m_H$ is from the $cg \to tH \to tt\bar t$ process,
while the slightly higher bound on $m_{H^+}$ arises from combining
$B_q$ mixing and $\bar bg \to \bar tH^+$ constraints 
and the opening of $H^+ \to AW^+$ decay.
Although not our benchmark,
{for $m_A = 400$ GeV, $m_{H}, m_{H^+} = 500, 530$ GeV
and $\rho_{tt}\sim 1.1$,  $\rho_{tc}\sim 0.9$,} 
%
{\it we find that \bhpm could reach up to 
$11\sigma$ significance with full Run~2 data!} 
While exciting 
if the excess is confirmed, it could also 
become problematic for the g2HDM if \bhpm is not seen.} 

{So far we have not considered the $\rho _{tu}$ coupling, which} 
can induce $ug \to b H^+ \to b t \bar b$, where 
the valence PDF means stronger constraint~{\cite{ttou}} 
from CRW and SR12 of Ref.~\cite{Sirunyan:2019wxt}, 
which weakens for larger $\rho_{tt}$. 
Taking $\rho_{tt} = 0.6$ and all other $\rho_{ij} = 0$, 
we find the CRW of Ref.~\cite{Sirunyan:2019wxt} excludes 
$|\rho_{tu}| \gtrsim 0.1\, (0.2)$ for BP1 (BP2) at 95\% C.L.,
with constraint from SR12 weaker. 
Taking $\rho _{tu} = 0.1$ 
 {(a rather large value in view of mass-mixing hierarchy protection)},
 $\rho_{tt}=0.6$, 
we find $\sim 3.2 \sigma$ ($2.7\sigma$) 
significance for BP1 (BP2) with full Run 2 data.
If $\rho_{tu}$-induced $bH^+$ production dominates, 
the valence PDF in $pp$ collisions would imply 
an asymmetry with $\bar b H^-$,
i.e. a charge asymmetry of $t \bar bb\bar b$ vs $\bar t bb\bar b$.
The effects from $bg\to \bar u H^+ \to \bar u t \bar b$ 
and $\bar t H^+ \to \bar t u \bar b$ are negligible.
Note that $B \to \mu \bar \nu$ 
decay probes~\cite{Hou:2019uxa} 
the $\rho_{tu}\rho_{\tau\mu}$ product at Belle~II.

One may have same-sign top signature 
via $cg\to tA/tH \to t t \bar c$. 
Following the same analysis of Refs.~\cite{Kohda:2017fkn,Hou:2018zmg},
we find BP1 may have $\sim3.5\sigma$ significance with full Run 2 data, 
but below $\sim1\sigma$ for BP2 due to dilution from $A/H\to t\bar t$ decay 
and falling parton luminosity.

Single-top studies may contain \bhpm events.
For $\rho_{tc} = 0.4$ and  $\rho_{tt} = 0.6$, 
we find the combined cross sections for $pp\to H^{+}[t\bar b] j$, 
$H^{+}[c\bar b] t$  
can contribute 15.2 (2.9)~pb for BP1 (BP2),
well within the $2\sigma$ error of current $t$-channel 
single-top~\cite{Aaboud:2016ymp,Sirunyan:2018rlu} measurements.
The situation is similar for Run 1 with $s$-channel single-top. 

We have not included uncertainties from scale dependence and PDF~\cite{Buza:1996wv,Maltoni:2012pa}, where the latter is 
sizable for processes initiated by heavy quarks. 
Using LO signal cross sections can also bring in some uncertainties,
e.g. higher order corrections~\cite{Kidonakis:2010ux,highorder}
to $\sigma(bg\to t H^+)$ may be $30$--$40\%$ for $m_{H^+} \sim 300$--500 GeV. 
A detailed study of such uncertainties is left for the future,
and is part of the reason why we adopt 
conservative $\rho_{tc}$, $\rho_{tt}$ values.

Finally, our 300--500 GeV mass range is not just for its promise. 
Significance can still be high at higher masses for 
larger $\rho_{tc}$, $\rho_{tt}$, 
but the decoupling $\mu_{22}^2$ would have to become larger~\cite{Hou:2017hiw} 
({as can be seen from $\mu_{22}^2/v^2 \simeq 3.78$ for}
BP2 in Table~\ref{bench}),
which would start to damp the EWBG motivation.
{But the $cg \to bH^+$ process can certainly be pursued 
for heavier $m_{H^+}$ at higher luminosities.}

In summary,
extra top Yukawa couplings $\rho_{tc}$ and $\rho_{tt}$ 
enter $\bar cbH^+$ and $\bar tbH^+$ couplings without CKM suppression,
leading to the $cg \to bH^+ \to bt\bar b$ signature of 
lepton plus missing energy and three $b$-jets.
For conservative $\rho_{tc},\, \rho_{tt} \sim 0.5$, 
evidence could already emerge with full LHC Run 2 data 
for $m_{H^+} = 300$--500 GeV, 
with discovery at 300 fb$^{-1}$ and beyond,
which would unequivocally point to physics beyond the Standard Model.
%

\vskip0.1cm
\noindent{\bf Acknowledgments} \
We thank Kai-Feng Chen for discussions,
and the support of MOST 106-2112-M-002-015-MY3 and 108-2811-M-002-537, 
and NTU 108L104019.

\vskip0.1cm
\noindent{\bf Note Added} \
After completion of this work, we noticed that the result 
for $pp\to \bar t b H^+ \to \bar t  b t \bar b$ search by ATLAS 
has been updated with full Run 2 dataset~\cite{ATLAS:2020jqj}. 
We have checked that the chosen $\rho_{tt}$ values for the BPs are still allowed 
by current data.


\end{document}